\documentstyle[preprint,eqsecnum,aps]{revtex}
\begin{document}
\draft
\preprint{HYUPT-96/3, SNUTP 96-070, cond-mat/9606165}
\title{Thermodynamic Properties of Generalized Exclusion Statistics}
\author{Hyun Seok Yang, Bum-Hoon Lee and Chanyong Park}
\address{Department of Physics, Hanyang University, Seoul 133-791, Korea}
\maketitle

\begin{abstract}
We analytically calculate some thermodynamic quantities of an ideal 
$g$-on gas obeying generalized exclusion statistics. 
We show that the specific heat of a $g$-on gas ($g \neq 0$) 
vanishes linearly in any dimension as $T \rightarrow 0$ 
when the particle number is conserved and exhibits 
an interesting dual symmetry that relates the particle-statistics 
at $g$ to the hole-statistics at $1/g$ at low temperatures. 
We derive the complete solution for the cluster coefficients $b_l(g)$ 
as a function of Haldane's 
statistical interaction $g$ in $D$ dimensions. 
We also find that the cluster coefficients $b_l(g)$ and 
the virial coefficients $a_l(g)$ are 
exactly mirror symmetric ($l$=odd) or 
antisymmetric ($l$=even) about $g=1/2$.
In two dimensions, we completely determine the closed forms 
about the cluster and 
the virial coefficients of the generalized 
exclusion statistics, which exactly agree with the virial 
coefficients of an anyon gas of linear energies. 
We show that the $g$-on gas with zero chemical potential 
shows thermodynamic properties similar to 
the photon statistics. 
We discuss some physical implications of our results.
  
\end{abstract}

\def\be{\begin{equation}}
\def\ee{\end{equation}}
\def\bea{\begin{eqnarray}}
\def\eea{\end{eqnarray}}
\def\ba{\begin{array}}
\def\ea{\end{array}}
\def\l{\label}
\def\r{\ref}
\def\c{\cite}
\narrowtext

\section{Introduction}
Recently, there has been extensive interest 
in generalized exclusion statistics (GES) 
\c{ha91,wu94,nw94,ra95,ms94,kn95,do94,jc94,lo94,is94,bm96,fk96,sb95,po96} 
initiated by Haldane. 
It has been realized \c{wu94,ms94,do94,jc94,lo94,is94,bm96,fk96} 
that there are several models obeying GES 
in which interparticle interactions can be 
regarded as purely statistical interactions by generalized exclusion 
principle. GES defined by Haldane generalizes the Pauli's 
exclusion principle through 
the linear differential relation \c{ha91}
\be
\l{g}
\Delta d_{\alpha}=-\sum_{\beta} g_{\alpha\beta}\Delta N_{\beta},
\ee
where $\Delta d_{\alpha}$ is the change of the available one-particle 
Hilbert-space dimension when the number of added particles amounts 
to $\Delta N_{\beta}$ and $\alpha$ and $\beta$ indicate different 
particle species. This definition of statistics is independent of 
the space dimension and obviously interpolates 
between boson $(g_{\alpha\beta}=0)$ and fermion 
$(g_{\alpha\beta}=\delta_{\alpha\beta})$ statistics continuously. 
Note $g_{\alpha\beta}$ can have arbitrary nonnegative values. 

An earlier form of fractional statistics-anyons \c{lm77} 
is related to braiding properties of particle trajectories, 
which is a peculiar 
property of two dimensions unlike GES. GES superficially seems 
to have little to do with the braid-group notion of 2$D$ statistics. 
However, several people have shown \cite{ha91,wu94,do94,jc94,lo94} 
that an anyon gas in the lowest Landau 
level (LLL) satisfies the GES given by the statistical interaction 
$g=\alpha$ (braiding statistics parameter). 
Moreover, Murthy and Shankar have argued \c{ms94} 
that anyons obey the GES by 
relating the statistical interaction $g$ to the high-temperature 
limit of the second virial coefficient (VC) of the anyon gas 
which shows a nontrivial dependence 
on 2$D$ braid statistics \c{as85}. 

As discussed by Nayak and Wilczek \c{nw94}, free anyons are not 
ideal $g$-ons but interacting $g$-ons. 
It will be an interesting question 
whether the ideal GES provides good approximations of the anyon 
model in which the Chern-Simons flux is attached 
to charged particles \c{lm77}. 
In this respect, it will be shown that ideal 
$g$-on statistics provides a good leading approximation 
of anyon statistics, i.e., the anyon statistics 
has a correct limit of GES, in the first-order 
perturbation-cluster expansion \c{se91,do92} and 
in the linear energies 
in a regulating harmonic potential \c{da96}. 

In this paper, we analytically calculate some thermodynamic quantities 
of an ideal $g$-on gas with $g_{\alpha\beta}=g\delta_{\alpha\beta}$. 
For the single-particle energy given by $\epsilon (p)=a p^n$, 
we verify that exclusion statistics satisfy 
the third law of thermodynamics as 
conjectured by Nayak and Wilczek \c{nw94}, i.e., 
when the particle number is conserved, 
the specific heat of a $g$-on gas ($g \neq 0$) vanishes linearly 
in any dimension as $T \rightarrow 0$ and show that 
the specific heat at low temperature exhibits 
an interesting dual symmetry 
that relates the particle-statistics 
at $g$ to the hole-statistics at $1/g$. 
We derive the complete solution for the cluster coefficients $b_l(g)$ 
and thus determine the VCs $a_l(g)$ as a function of 
Haldane's statistical interaction $g$ 
in $D$ dimensions. 
We also demonstrate that the cluster coefficient (CC) $b_l(g)$ and 
the VC $a_l(g)$ are 
exactly mirror symmetric ($l$=odd) or 
antisymmetric ($l$=even) about $g=1/2$, so that the semion $(g=1/2)$
which is the case of 1$D$ spinon as discussed 
by Haldane \c{ha91,has} must exhibit a peculiar property, 
$a_{2l}(\frac{1}{2})=b_{2l}(\frac{1}{2})=0$ 
for $\forall l \in {\bf N}$. 
In the case of $D=n$, we completely determine the cluster 
and the virial coefficients of GES in closed forms, 
and the results exactly agree with the VCs 
of anyon gas obtained by Sen \c{se91} and 
Dasni\`eres de Veigy \c{da96}. 
This result implies that a free anyon gas with 
a small $\alpha$ or with linear energies 
can be well described by an ideal $g$-on gas. 
We also show that the $g$-on gas with zero chemical potential 
shows thermodynamic properties similar to 
the photon statistics 
and all the thermodynamic quantities for a given volume 
depend only on the temperature and 
their dependences can be determined by simple 
dimensional arguments. 
Finally, we discuss some physical implications of our results.

\section{Thermodynamics of generalized exclusion statistics} 
Consider a one-particle energy spectrum divided 
into a large number of energy cells $\epsilon_{\alpha}$, 
each of which contains $G_{\alpha}$ independent levels 
and $N_{\alpha}$ identical particles. 
In the limit of a discrete energy spectrum, 
$G_{\alpha}$ would be the degeneracy factor. 
If $d_{N_{\alpha}}$ is the dimension of the one-particle 
Hilbert space in the $\alpha$th cell with 
the coordinates of $N_{\alpha}-1$ particles held fixed, 
Haldane's definition of exclusion principle, Eq.(\r{g}), 
leads to
\[d_{N_{\alpha}}=G_{\alpha}-g(N_{\alpha}-1).\] 
Let $W(\{N_{\alpha}\})$ be the number of configurations of 
the system corresponding to the set of occupation numbers 
$\{N_{\alpha}\}$. Then an elementary combinatorial argument 
gives \c{wu94,nw94,jc94}
\be
\l{wn}
W(\{N_{\alpha}\})=\prod_{\alpha}\frac{(d_{N_{\alpha}}+N_{\alpha}-1)!}
{N_{\alpha}!(d_{N_{\alpha}}-1)!}.
\ee
Under the constraints of fixed particle number and energy,
\be
\l{cen}
\ba{l}
N=\sum_{\alpha} N_{\alpha},\\
E=\sum_{\alpha} \epsilon_{\alpha} N_{\alpha},
\ea
\ee
the grand partition function $Z$ is determined by 
Haldane-Wu state-counting rule (\r{wn}) as follows \c{..}
\be
\l{z}
Z=\sum_{\{N_{\alpha}\}}W(\{N_{\alpha}\})
\mbox{exp}\{\sum_{\alpha}N_{\alpha}(\mu-\epsilon_{\alpha})/kT\},
\ee
where $\mu$ and $T$ are the Lagrange multipliers 
incorporating the constraints, Eq. (\r{cen}), of 
fixed particle number and energy, respectively. 

The stationary condition of the grand partition function $Z$ 
with respect to $N_{\alpha}$ gives the statistical distribution
$n_{\alpha}\equiv {\bar N}_{\alpha}/G_{\alpha}$ of an
ideal $g$-on gas with chemical potential $\mu$ and temperature $T$ 
as derived by Wu \c{wu94}:
\be
\l{dn}
n_{\alpha}=\frac{1}{w({\zeta_{\alpha}})+g},
\ee
where the function $w({\zeta}_{\alpha})$ satisfies 
the functional equation
\be
\l{aew}
w({\zeta}_{\alpha})^g [1+w({\zeta}_{\alpha})]^{1-g}
= \zeta_{\alpha} \equiv e^{(\epsilon_{\alpha}-\mu)/kT}.
\ee
As with usual statistics, in the thermodynamic limit, all the thermodynamic functions 
of GES can be evaluated around the set 
$\{{\bar N}_{\alpha}\}$ of the most probable occupation numbers.

The Eqs.(\r{dn}) and (\r{aew}) give correct solutions 
for the familiar Bose $(g=0)$ and Fermi $(g=1)$ distributions 
because $w({\zeta}_{\alpha})={\zeta}_{\alpha}-1$ for $g=0$ 
and $w({\zeta}_{\alpha})={\zeta}_{\alpha}$ for $g=1$. 
Since the function ${\cal F}_g(w)\equiv w^g(1+w)^{1-g}-\zeta_{\alpha}$ 
is a monotonically increasing function for $w>0$, 
the algebraic equation (\r{aew}) has 
only one positive root for $g \neq 0$ and $T \neq 0$ 
and thus the statistics is unique for a given $g$. 
Also, $n_{\alpha}$ decreases monotonically as $g$ increases. 
In other wards, as $g$ is increased from boson $(g=0)$ toward 
fermion $(g=1)$ statistics, the occupation number $n_{\alpha}$ decreases 
for a fixed ${\zeta}_{\alpha}$. 
 
At $T=0$, the ditribution (\r{dn}) reveals a quite surprising 
phenomenon, such as {\it formation of Fermi surface} 
for $g\neq 0$ \c{wu94,nw94}:
\be
\l{fs}
n_{\alpha}=\left\{ \ba{ll}
                    1/g, & \mbox{if $\epsilon_{\alpha}
                                     < \mu=\epsilon_F$,} \\
                    0,   & \mbox{if $\epsilon_{\alpha}
                                      > \mu=\epsilon_F$,}
                   \ea
           \right.
\ee
where the Fermi surface $\epsilon_{\alpha}=\epsilon_F$ is determined 
by the condition $\sum_{\epsilon_{\alpha} < \epsilon_F}
G_{\alpha}n_{\alpha}=N$. 
This property has important effects, to be demonstrated later, on the low-temperature 
thermodynamics of the $g$-on gas.
Since the statistics at $g=0$ is the Bose-Einstein statistics and is well-known, 
we will not present a detailed analysis in this paper.

As observed by Nayak and Wilczek \c{nw94}
and Rajagopal \c{ra95}, 
Eq. (\r{aew}) displays a duality property 
that relates the particle-statistics at $g$ 
to the hole-statistics at $1/g$ 
with a rescaled ${\zeta}_{\alpha}^{-\frac{1}{g}}$:
\be
\l{d}
w({\zeta}_{\alpha}^{-\frac{1}{g}};\frac{1}{g})=1/w({\zeta}_{\alpha};g).
\ee

In terms of the distribution $n_{\alpha}$ in Eq.(\r{dn}), 
the thermodynamic potential 
$\Omega=-kT \ln Z$ is given by
\be
\l{pv}
\Omega \equiv -PV = -kT \sum_{\alpha}G_{\alpha} 
\ln \frac{1+(1-g)n_{\alpha}}{1-g n_{\alpha}}.
\ee

>From the Eq.(\r{aew}), one can see the following 
properties on $w({\zeta}_{\alpha})$:
\be
\l{winf}
\mbox{If}\;\epsilon_{\alpha}\rightarrow\infty,\;\;
\mbox{then}\;w({\zeta}_{\alpha})\rightarrow 
\zeta_{\alpha}=e^{(\epsilon_{\alpha}-\mu)/kT}
=\frac{1}{z}e^{\epsilon_{\alpha}/kT},
\ee
and,
\be
\l{w0}
\mbox{if}\;\epsilon_{\alpha}\rightarrow0 \; \mbox{and} 
\; T \rightarrow 0,\;\;\mbox{then}\; w({\zeta}_{\alpha}) 
\rightarrow  w_0 \equiv \left\{ \ba{ll}
                    e^{-\epsilon_F/gkT}, & \mbox{if $g\neq0$}, \\
                    \frac{1-z}{z},   & \mbox{if $g=0$},
                   \ea
           \right. 
\ee
where the fugacity $z \equiv e^{\mu/kT}$.

In our calculations, we will use periodic boundary conditions 
over the system's volume $V\equiv L^D$; thus, the $D$-dimensional momentum 
eigenvalue ${\bf p}$ of 
single-particle is given by
\[{\bf p}=\frac{2\pi\hbar}{L}{\bf n}\]
where the $D$-dimensional vector ${\bf n}$ takes only integer values. 
Then, in thermodynamic limit as $V\rightarrow \infty$, 
a sum over ${\bf p}$ can be replaced by an integration:
\[\sum_{{\bf p}} \rightarrow \frac{V}{(2\pi\hbar)^D}\int d^D {\bf p}.\]

A state of an ideal $g$-on gas can be specified by specifying a set 
of occupation numbers $\{n_{\bf p}\}$ so defined that 
there are $n_{\bf p}$ particles having the momentum ${\bf p}$ 
in the state under consideration. 
Of course, as pointed out by Nayak and Wilczek \c{nw94}, 
we must take the cell size of the single-particle 
energy spectrum to be sufficiently large, 
i.e. $G_{\alpha}\gg 1$, in order to safely ignore 
the specific complication of GES coming from negative probabilities. 
Then the occupation number $N_{\alpha}$ of 
the $\alpha$th cell will be the sum of $n_{\bf p}$ 
over all the levels in the $\alpha$th cell. 
With this in mind, the total energy $E$ and 
the total number of particles $N$ in Eq.(\r{cen}) 
can be represented as the sum over the momentum ${\bf p}$: 
\bea
\l{gn}
&& N=\sum_{{\bf p}} {\bar n}_{{\bf p}}(g),\\
\l{ge}
&&E =\sum_{{\bf p}} \epsilon_{{\bf p}} {\bar n}_{{\bf p}}(g)
\eea
with ${\bar n}_{{\bf p}}(g)={\bar N}_{\alpha}/G_{\alpha}$ 
determined by Eqs.(\r{dn}) and (\r{aew}). 
In the following, we will denote the statistical distribution 
${\bar n}_{{\bf p}}(g)$ at statistics $g$ as $n_g$ for simplicity.
We assume the single-particle energy of an ideal $g$-on gas is given by
\be
\l{e}
\epsilon_{{\bf p}}=a p^n,\;\;\;p=|{\bf p}|.
\ee 

Using spherical coordinates in $D$ dimensions, 
the pressure $P$ given by Eq.(\r{pv}) can be represented as 
the following integration in terms of the statistical 
distribution $n_g$ at $g$:
\be
\l{p}
\ba{ll}
\frac{P}{kT}&= \frac{2 \pi^{D/2}a^{-D/n}}
{(2\pi\hbar)^D \Gamma(D/2)n}\int^{\infty}_{0}d\epsilon\; 
\epsilon^{\frac{D}{n}-1}\ln \frac{1+(1-g)n_g}{1-g n_g},\\
&= \frac{2 \pi^{D/2}a^{-D/n}}{(2\pi\hbar)^D \Gamma(D/2)D}kT 
\int^{\infty}_{0}d\epsilon\; \epsilon^{D/n} n_g,
\ea
\ee
where the last step is obtained through a partial integration 
and Eq.(\r{dn}). 
In the same way, the particle density $\rho\equiv N/V$ is 
given by 
\be
\l{nh}
\ba{ll}
\rho&=\frac{2 \pi^{D/2}a^{-D/n}}{(2\pi\hbar)^D \Gamma(D/2)n}
\int^{\infty}_{0}d\epsilon \;\epsilon^{\frac{D}{n}-1}n_g\\
&=\frac{z}{kT}\frac{\partial P}{\partial z},
\ea
\ee
where the second step is obtained through a partial integration 
and by using the relation $kT\frac{dn_g}{d\epsilon}=-z\frac{dn_g}{dz}$. 

Equations (\r{p}) and (\r{nh}) can 
be written in the following standard form \c{..}
\bea
\l{p1}
&&\frac{P}{kT}\equiv\frac{1}{\lambda^D}f_{\frac{D}{n}+1}^{(g)}(z)\\
\l{n1}
&&\rho\equiv\frac{1}{\lambda^D}f_{\frac{D}{n}}^{(g)}(z),
\eea
where
\be
\l{fz}
f_{\nu}^{(g)}(z)=\frac{1}{\Gamma(\nu)}
\int_{0}^{\infty}dy\;y^{\nu-1}n_g(z^{-1}e^y),
\;\;y=\frac{\epsilon}{kT}
\ee
and the thermal wavelength $\lambda^D\equiv 
n\frac{(2\pi\hbar)^D a^{\frac{D}{n}}}
{2\pi^{\frac{D}{2}}(kT)^{\frac{D}{n}}}
\frac{\Gamma(D/2)}{\Gamma(D/n)}$. 
By the Eq.(\r{n1}), one can determine the Fermi energy $\epsilon_F$
in Eq.(\r{fs}) in terms of the density $\rho$:
\be
\l{fe}
\epsilon_F=[g(4\pi)^{\frac{D}{2}}
\Gamma(\frac{D}{2}+1)]^{\frac{n}{D}}a(\hbar\rho^{\frac{1}{D}})^n.
\ee
>From the definition in the Eq.(\r{fz}), 
we obtain the relation
\be
\l{pnr}
f_{\frac{D}{n}}^{(g)}(z)=z\frac{\partial}{\partial z}
f_{\frac{D}{n}+1}^{(g)}(z).
\ee
 
>From Eq.(\r{ge}), the energy $E$ can be also expressed as  
\be
\l{eh}
\frac{E}{V}=\frac{D}{n}\frac{kT}{\lambda^D}f_{\frac{D}{n}+1}^{(g)}(z),
\ee
which obviously exhibits the statistics-independent relation
\be
\l{per}
P=\frac{n}{D}\frac{E}{V}.
\ee

According to the fundamental principles of statistical mechanics, 
the entropy $S$ of the system is given by
\bea
\l{s}
\frac{S}{k}&=&\sum_{\bf p}\left\{
n_g\frac{\epsilon_{\bf p}-\mu}{kT}+
\ln \frac{1+(1-g)n_g}{1-g n_g}\right\}\nonumber\\
&=&\frac{D+n}{n}\frac{PV}{kT}-N\mbox{ln}z,
\eea
where the second expression is obtained by using Eq. (\r{per}). 
The density fluctuations in 
the grand partition function $Z$ can 
be expressed as
\bea
\l{df}
<N^2>-<N>^2&=&z\frac{\partial}{\partial z}
z\frac{\partial}{\partial z}\mbox{ln}Z=
\frac{V}{kT}z\frac{\partial}{\partial z}
z\frac{\partial}{\partial z}P\nonumber\\
&=&{\bar N}kT\rho \kappa_T,
\eea
where the isothermal compressibility $\kappa_T \equiv -\frac{1}{V}
(\frac{\partial V}{\partial P})_T=\frac{1}{\rho}
(\frac{\partial \rho}{\partial P})_T$ and Eq. (\r{nh}) is used. 
It can be easily shown that the isothermal compressibility 
in the case of $\mu \neq 0$ is given by
\be
\l{cp}
\kappa_T=\frac{1}{kT}\frac{1}{\lambda^D\rho^2}
z\frac{\partial}{\partial z} f^{(g)}_{\frac{D}{n}}(z).
\ee
Equation (\r{df}) shows that the density fluctuations 
of a $g$-on gas 
are vanishingly small in the thermodynamic limit, 
provided that $\kappa_T$ is finite. 

\section{General properties of generalized exclusion statistics}
\subsection{General cases}

First, we will calculate the thermodynamic quantities, such as 
the total energy $E$, 
the specific heat $C_V\equiv(\partial E/\partial T)_V$ 
at constant volume, the entropy $S$, and the isothermal 
compressibility $\kappa_T$, 
for the single-particle energies given by 
$\epsilon (p)=a p^n$ at low temperature and high density, 
i.e., $\lambda^D \rho \gg 1$. 

At the low-$T$ and high-$\rho$ limit $(g\neq 0)$, 
the Eq.(\r{fz}) can be evaluated by the so-called 
Sommerfeld expansion method \c{..} 
after partial integrations with respect to $y$:
\be
\l{pfz}
f_{\nu}^{(g)}(z)=-\frac{1}{\Gamma(\nu+1)}
(\frac{\mu}{kT})^{\nu}
\sum_{j=0}\left(\ba{c} \nu \\ j
      \ea  \right)(\frac{kT}{\mu})^j C_j^g,
\ee
where the numerical numbers $C_j^g$ at statistics $g$ is defined by
\be
C_j^g=\int^{\infty}_{-\frac{\mu}{kT}}dx\; x^j
\frac{dn_g(x)}{dx},\;\;x=(\epsilon-\mu)/kT
\ee
and $C_0^g=-\frac{1}{g}$. 
Then Eqs.(\r{n1}) and (\r{eh}) are given by
\bea
\l{nd}
&&\frac{N}{V}=-\frac{1}{\lambda^D(kT)^{\frac{D}{n}}\Gamma(D/n)}
\frac{n}{D}\mu^{D/n}\sum_{j=0}
\left(\ba{c} D/n \\ j
      \ea  \right)(\frac{kT}{\mu})^j C_j^g,\\
\l{ed}
&&\frac{E}{V}=-\frac{1}{\lambda^D(kT)^{\frac{D}{n}}\Gamma(D/n)}
\frac{n}{D+n} \mu^{1+\frac{D}{n}}\sum_{j=0}
\left(\ba{c} 1+\frac{D}{n} \\ j
      \ea  \right)(\frac{kT}{\mu})^j C_j^g.
\eea
Let us rewrite Eqs.(\r{nd}) and (\r{ed}) 
in the following forms \c{nw94}
\bea
\l{nD}
&&\epsilon_F^{D/n}=\mu^{D/n}\left[1-g\sum_{j=1}(\frac{kT}{\mu})^j
\left(\ba{c} D/n \\ j
      \ea  \right) C_j^g \right],\\
\l{eD}
&&E=E_0(\frac{\mu}{\epsilon_F})^{1+\frac{D}{n}}\left[1-g\sum_{j=1}
(\frac{kT}{\mu})^j
\left(\ba{c} 1+\frac{D}{n} \\ j
      \ea  \right) C_j^g \right],
\eea
where the Fermi energy $\epsilon_F$ is given by Eq.(\r{fe})  
and $E_0=\frac{D}{D+n}N\epsilon_F$ which is the ground state 
energy of the $g$-on gas 
at the given density.
 
Here, we will convert the $x$-integration 
into a $w$-integration using Eqs.(\r{dn}) and (\r{aew}):
\bea
\l{cin}
C_j^g&=&\int^{\infty}_{-\frac{\mu}{kT}}dx\; 
      x^j \frac{dn_g(x)}{dx} \nonumber\\  
   &=&-\int^{\infty}_{0}dw\; \frac{1}{(w+g)^2} 
     \{g\ln w +(1-g)\ln(1+w)\}^j \nonumber\\
   && +\int^{w_0}_{0}dw\; \frac{1}{(w+g)^2} 
     \{g\ln w +(1-g)\ln(1+w)\}^j \nonumber\\
   &=&-\int^{\infty}_{0}dw\; \frac{1}{(w+g)^2} 
     \{g\ln w +(1-g)\ln(1+w)\}^j +O(e^{-\epsilon_F/gkT}).
\eea
Since $w_0\simeq e^{-\epsilon_F/gkT}\ll 1$, the second integration 
can be certainly neglected, which is the reason we use the 
partial-integrated forms for $N/V$ and $E$. 
The dual symmetry given by Eq.(\r{d}) relates the coefficients 
$C_j^g$ at statistics $g$ to the coefficients 
$C_j^{\frac{1}{g}}$ at $\frac{1}{g}$ 
up to $O(e^{-\epsilon_F/gkT})$ as follows:
\be
\l{cd}
C_j^g=(-1)^j g^{j-2} C_j^{\frac{1}{g}},
\ee
which can be proved from Eq.(\r{cin}) by the substitution, 
$w \rightarrow \frac{1}{w}$.
Note that only fermions $(g=1)$ have the {\it exact} particle-hole 
dual symmetry: $C_j^1=0$ for all odd $j$ \c{..}. 
After a partial integration, $C_j^g\; (j \geq 1)$ becomes
\be
C_j^g=- g^{j-1}\{\ln w\}^j|_{w=0}- j\int^{\infty}_{0}dw\; 
     \frac{1}{w(1+w)} 
     \{g\ln w +(1-g)\ln(1+w)\}^{j-1}.
\ee
Using the integral formulas 
\bea
\l{z2}
&&\int dw\;\frac{\ln w}{w(1+w)}=\frac{1}{2}(\ln\frac{w}{1+w})^2+
\int dw\;\frac{\ln(1+w)}{w(1+w)}\\
\l{ze2}
&&\int^{\infty}_{0}dw\;\frac{\ln(1+w)}{w(1+w)}
=\zeta(2)=\frac{\pi^2}{6},
\eea
we can easily calculate $C_1^g$ and $C_2^g$. The results are
\be
\ba{l}
C_1^g=0,\\
C_2^g=-\frac{\pi^2}{3}.
\ea
\ee
The higher coefficients of $C_j^g$ can be obtained numerically
\be
\l{Cc}
\ba{llll}
C^{\frac{1}{2}}_3=-3.6062, & C^2_3=7.2123, 
& C^{\frac{1}{2}}_4=-29.2227, & C^2_4=-116.891 \\
C^{\frac{1}{3}}_3=-4.8082, & C^3_3=14.4247,
& C^{\frac{1}{3}}_4=-26.6973, & C^3_4=-240.276 \\
C^{\frac{1}{4}}_3=-5.4093, & C^4_3=21.637, 
& C^{\frac{1}{4}}_4=-25.9758, & C^4_4=-415.612,\;\mbox{etc}.
\ea
\ee
The above results clearly display the dual symmetry in Eq. (\r{cd}). 

The chemical potential $\mu$, the total energy $E$, and 
the specific heat $C_V$ are given by
\bea
\l{mu}
&&\mu=\epsilon_F\left[1-g\frac{\pi^2}{6}(\frac{D}{n}-1)
(\frac{kT}{\epsilon_F})^2+O(T^3)\right]\\
\l{le}
&&E=E_0\left[1+g\frac{\pi^2}{6}(1+\frac{D}{n})
(\frac{kT}{\epsilon_F})^2+O(T^3)\right]\\
&&C_V/k = g\frac{\pi^2}{3}(1+\frac{D}{n})
\frac{E_0}{\epsilon_F^2}kT+O(T^2).
\eea
Note that, at low temperatures, when the temperature increases,
the chemical potential $\mu$ increases in the case of $D/n<1$, 
remains stable in $D/n=1$, and decreases in the case of $D/n>1$.  
This is consistent with the well-known results for a fermion gas 
when $n=2$ \c{ck}. 
Using Eqs. (\r{mu}) and (\r{le}), one can easily calculate 
the entropy $S$ defined by Eq.(\r{s}):
\bea
\l{ls}
\frac{S}{k}&=&\frac{1}{kT}(\frac{D+n}{D}E-\mu N)\nonumber\\
&=&g\frac{\pi^2}{3}(1+\frac{D}{n})
\frac{E_0}{\epsilon_F^2}kT+O(T^2)\\
&\simeq&C_V/k. \nonumber
\eea

When the particle number is conserved, 
the specific heat of a $g$-on gas ($g\neq 0$) vanishes 
linearly in any dimension as $T \rightarrow 0$. 
Therefore, as conjectured by Nayak and Wilczek \c{nw94},
GES obviously satisfies the third law of thermodynamics, which implies that 
the ground state of a many-body $g$-on system is nondegenerate. 
Our above results also 
agree with those of Nayak and Wilczek for the case of 
$g=\frac{1}{2}$ and, in the case of $g=1$ and $D/n=3/2$, 
reproduce the well-known results for a fermion gas \c{..}.     
The linear dependences on $T$ and $g$ of the specific heat of 
a $g$-on gas can be roughly 
understood in the same way for an ordinary fermion gas. 
Of course, the key point is the existence of 
Fermi surface at low temperature. 
(The energy of particles excited over 
$\epsilon_F$ is of order $gkT$, and the number of excited particles 
is of order $(kT/\epsilon_F)N$. Thus the total 
excitation energy amounts to $g(kT/\epsilon_F)NkT$, 
so that $C_V\approx g(kT/\epsilon_F)Nk$.) 
Note that the linear dependences on $T$ and $g$ of specific heat 
are satisfied even in D dimensions with $\epsilon (p)=a p^n$. 
Also the isothermal compressibility $\kappa_T$ given by 
Eq.(\r{cp}) can be calculated by using Eqs. (\r{nd}) and (\r{mu})
\be
\kappa_T=g\frac{D}{n}\Gamma(\frac{D}{2}+1)
(4\pi\hbar^2)^{\frac{D}{2}}\frac{a^{\frac{D}{n}}}
{\epsilon_F^{\frac{D}{n}+1}}\left[1-g\frac{\pi^2}{6}(\frac{D}{n}-1)
(\frac{kT}{\epsilon_F})^2+O(T^3)\right].
\ee
Note that the behavior of the leading term is linear in $g$ and 
the correction of order $T^2$ vanishes when $D=n$. 
   
Since the statistics at $g=0$ is a well-known bosonic one, 
we will present only the results for the statistics at $g=0$ for comparision:
\bea
\l{bn}
\frac{N}{V}&=&\frac{1}{\lambda^D}
\sum_{l=1}^{\infty}\frac{z^l}{l^{\frac{D}{n}}}
+\frac{1}{V}\frac{z}{1-z},\\
\l{be}
\frac{E}{V}&=&\frac{D}{n}\frac{kT}{\lambda^D}
\sum_{l=1}^{\infty}\frac{z^l}{l^{1+\frac{D}{n}}}.
\eea
When $T\rightarrow 0$ or $z \rightarrow 1$, 
the above results exhibit the characteristic properties 
of Bose-Einstein condensation \c{..}, 
unlike the statistics at $g\neq 0$. 
The important point is that only the Bose-Einstein statistics 
($g=0$) exhibits Bose-Einstein condensation $(D \ge 3)$.

Now, we will investigate the thermodynamic propreties 
at high $T$ and low $\rho$ in $D$ dimensions. 
This region corresponds to $\lambda^D \rho \ll 1$ 
or $z \ll 1$. For this purpose, we will take the 
following expansion of 
the function $f_{\nu}^{(g)}(z)$ in Eq.(\r{fz}):
\be
\l{fze}
f_{\nu}^{(g)}(z)=\sum_{l=1}^{\infty}b_l^{(\nu)}(g)z^l.
\ee
According to the statistical relation (\r{pnr}), 
the coefficients $b_l^{(\nu)}(g)$ satisfy the following relation
\be
\l{ccl}
b_l^{(\nu-1)}(g)=l\cdot b_l^{(\nu)}(g).
\ee 
The following results can be obtained from Eq.(\r{fze}): 
\bea
\l{vp1}
&&\frac{P}{kT}=\frac{1}{\lambda^D}f_{\frac{D}{n}+1}^{(g)}(z)
=\frac{1}{\lambda^D}\sum_{l=1}^{\infty}
b_l(g)\;z^l \\
\l{vn1}
&&\rho=\frac{1}{\lambda^D}f_{\frac{D}{n}}^{(g)}(z)
=\frac{1}{\lambda^D}\sum_{l=1}^{\infty}
l\;b_l(g)\;z^l,
\eea
where $b_l(g)\equiv b_l^{(\frac{D}{n}+1)}(g)$ 
is the $l$th cluster coefficient. 
The virial expansion \c{..} of the equation of state 
is defined to be
\be
\l{ve}
\frac{P}{kT}=\rho \sum_{l=1}^{\infty}
a_l(g)({\lambda^D}\rho)^{l-1},
\ee
where $a_l(g)$ is called the $l$th VC.
By substituting Eq.(\r{ve}) into Eqs.(\r{vp1}) and (\r{vn1}) 
and requiring that the resulting equation to be satisfied 
for every $z$, we obtain
\be
\l{vcs}
\ba{ll}
a_1 =&b_1=1 \\
a_2 =&-b_2 \\
a_3 =&4b_2^2-2b_3 \\
a_4 =&-20b_2^3+18b_2b_3-3b_4 \\
a_5 =&2(56b_2^4-72b_2^2b_3+9b_3^2+16b_2b_4-2b_5) \\
a_6 =&-672b_2^5+1120b_2^3b_3-315b_2b_3^2-280b_2^2b_4 \\
     & +60b_3b_4+50b_2b_5-5b_6 \\
 \cdots.&
\ea
\ee

To get the $l$th CCs $b_l(g)$, let us expand 
the distribution $n(\eta)$ as
\be
\l{nze}
n(\eta)=\sum_{l=1}^{\infty}c_l(g)\eta^l,
\ee
where $\eta=\zeta^{-1}=ze^{-\frac{\epsilon}{kT}}$. 
Then the coefficients $b_l^{(\nu)}(g)$ in Eq.(\r{fze}) are 
\be
\l{cc}
b_l^{(\nu)}(g)=\frac{c_l(g)}{l^{\nu}}.
\ee
It can be easily shown that 
for bosons $c_l(0)=1$ and for fermions 
$c_l(1)=(-1)^{l+1}$. 
Although the coefficients $b_l^{(\nu)}(g)$ or $c_l(g)$
for general $g$ can be 
obtained by expanding the Eqs.(\r{dn}) 
and (\r{aew}) in terms of $\zeta^{-1}$, 
we will show in the next section the CC $b_l^{(\nu)}(g)$  
can be expressed by the closed form
\be
\l{cg}
b_l^{(\nu)}(g)=\frac{1}{l^{\nu}}\prod_{m=1}^{l-1}\frac{m-gl}{m}.
\ee
All the VCs, $a_l(g)$, can be obtained from the Eq.(\r{vcs}) 
because we have already known all the CCs, $b_l(g)$, from Eq.(\r{cg}). 

The first few terms of the pressure $P$ and 
the particle density $\rho$ in Eqs.(\r{vp1}) and (\r{vn1}), respectively, 
are given by
\bea
\l{hvp}
&&P=\frac{kT}{\lambda^D}(z-\frac{2g-1}{2^{\frac{D}{n}+1}}z^2
  +\frac{9g^2-9g+2}{2\cdot3^{\frac{D}{n}+1}}z^3 +\cdots),\\
\l{hvn}
&&\lambda^D \rho= z-\frac{2g-1}{2^{\frac{D}{n}}}z^2
  +\frac{9g^2-9g+2}{2\cdot3^{\frac{D}{n}}}z^3+\cdots.
\eea
The virial expansion (\r{ve}) becomes
\be
PV = NkT\left[1+\frac{2g-1}{2^{1+\frac{D}{n}}}\lambda^D\rho
+\left\{\frac{(2g-1)^2}{4^{D/n}}-\frac{9g^2-9g+2}
{3^{1+\frac{D}{n}}}\right\}(\lambda^D\rho)^2+\cdots\right].
\ee  
In other words, the second VC and the third VC are
\bea
\l{a2}
&& a_2=\frac{2g-1}{2^{1+\frac{D}{n}}},\\
\l{a3}
&& a_3=\frac{(2g-1)^2}{4^{D/n}}-\frac{9g^2-9g+2}{3^{1+\frac{D}{n}}}.
\eea
Note that $a_3$, as well as $a_2$, is statistics-dependent 
in $D$ dimensions; $a_3$ is 
statistics-independent only in 2$D$, 
or more accurately in the case of $D/n=1$. 
Of course, when $g=0$ or $1$, our above results reproduce 
the standard results of statistical mechanics. 

Using the results given by Eqs. (\r{hvn}) and (\r{hvp}), 
the entropy $S$ in Eq.(\r{s}) and 
the compressibility $\kappa_T$ in Eq.(\r{cp}) 
in the Boltzmann limit are
\bea
\l{hs}
\frac{S}{k}
&\simeq& N[\frac{D+n}{n}-\mbox{ln}\lambda^D\rho]\nonumber\\
&&+N[\frac{D+n}{n}\frac{2g-1}{2^{1+\frac{D}{n}}}\lambda^D\rho
-\mbox{ln}(1+\frac{2g-1}{2^{\frac{D}{n}}}\lambda^D\rho)]\\
\l{hc}
\kappa_T&\simeq& \frac{1}{P}[1+\{\frac{9g^2-9g+2}
{2\cdot3^{\frac{D}{n}}}-\frac{(2g-1)^2}{4^{\frac{D}{n}}}\}
\frac{\lambda^{2D}}{(kT)^2}P^2].
\eea 
The first part in Eq.(\r{hs}) is just 
the Sackur-Tetrode equation \c{..}, 
and the second part is a correction due to quantum statistics. 
Note that this quantum correction has no linear dependence 
on $\rho$ when $D=n$ and that a sign change exists at $g=1/2$.

We will prove a very interesting mirror symmtery for 
the CC $b_l(g)$ and the VC 
$a_l(g)$ about $g=1/2$. 
Let us start with the following relation derived 
by the Eqs. (\r{dn}) and (\r{aew}):
\be
\l{aen}
(1-gn_g)^g [1+(1-g)n_g]^{1-g}=n_g\frac{1}{z}e^{\frac{\epsilon}{kT}}.
\ee
Now, Eq.(\r{aen}) can be rewritten as
\be
\l{man}
[1-(\frac{1}{2}+\tilde{g})n_{\tilde{g}}]^{\frac{1}{2}+\tilde{g}}
[1+(\frac{1}{2}-\tilde{g})n_{\tilde{g}}]^{\frac{1}{2}-\tilde{g}}
=n_{\tilde{g}}\frac{1}{z}e^{\frac{\epsilon}{kT}},
\ee
where $\tilde{g}=g-\frac{1}{2}$. 
Oviously, one can see that the form of Eq.(\r{man}) is invariant 
under the mirror transformation
\be
\ba{l}
\tilde{g} \rightarrow -\tilde{g}\\
z \rightarrow -z\\
n_{\tilde{g}} \rightarrow -n_{-\tilde{g}}.
\ea
\ee
The above mirror symmetry implies that, if $n_{\tilde{g}}(z)$ is a 
solution of Eq.(\r{man}), then $-n_{-\tilde{g}}(-z)$ is 
also a solution. Since the algebraic equation (\r{man}) 
must have a unique 
nonnegative solution for a given $g$ as remarked at the 
beginning of Sec.II, we can obtain the interesting 
symmetry property
\be
\l{mn}
n_{-\tilde{g}}(-z)=-n_{\tilde{g}}(z).
\ee
Note that the above mirror symmetry about the distribution $n_g$ 
is generally satisfied for all temperatures and densities. 

The following mirror symmetry is the direct consequence of 
Eq.(\r{mn}) derived from Eq.(\r{nze}):
\be
\l{cds}
\ba{l}
c_{2l+1}(-\tilde g)=c_{2l+1}(\tilde g)\\
c_{2l}(-\tilde g)=-c_{2l}(\tilde g).
\ea
\ee
The above mirror symmetry immediately leads to the mirror 
symmetry about the pressure $P$ and the particle density $\rho$
\be
\l{mpn}
\ba{l}
f_{D/n}^{(-\tilde g)}(-z)=-f_{D/n}^{(\tilde g)}(z)\;\;
\mbox{or}\;\;\rho(-z, -\tilde g)=-\rho(z, \tilde g)\\
f_{\frac{D}{n}+1}^{(-\tilde g)}(-z)=
-f_{\frac{D}{n}+1}^{(\tilde g)}(z)\;\;
\mbox{or}\;\;P(-z, -\tilde g)=-P(z, \tilde g).
\ea
\ee
>From the Eq.(\r{mpn}), one can easily verify the following 
mirror symmtery about $g=1/2$ of the CC $b_l(g)$ 
and the VC $a_l(g)$ in Eqs. (\r{vp1})-(\r{ve}):
\be
\l{mcv}
\ba{ll}
b_{2l+1}(-\tilde g)=b_{2l+1}(\tilde g),\;\;\;
& a_{2l+1}(-\tilde g)=a_{2l+1}(\tilde g),\\
b_{2l}(-\tilde g)=-b_{2l}(\tilde g),\;\;\;
&a_{2l}(-\tilde g)=-a_{2l}(\tilde g).
\ea
\ee
In particular, semions with $g=1/2$ exhibit a surprising property, 
$a_{2l}(\frac{1}{2})=b_{2l}(\frac{1}{2})=0$ 
for $\forall l \in {\bf N}$.
We can observe that our previous explicit results consistently show 
the mirror symmetries given by Eqs. (\r{mn})-(\r{mcv}). 

\subsection{D=n}

In this section, we will consider the special case of $D=n$, 
which allows more analytic results. This includes the 
usual non-relativistic system with 
$\epsilon_{{\bf p}}=p^2/2m$ in two dimensions 
as a special case. 
In the case of $D=n$, we have an explicit solution \c{wu94} 
directly obtained from Eq.(\r{n1}) and Eq.(\r{aew}):
\be
\l{wu}
\rho=\frac{1}{\lambda^D}\mbox{ln}\frac{1+w(z^{-1})}{w(z^{-1})}
\Rightarrow ze^{-g\lambda^D\rho}+e^{-\lambda^D\rho}=1.
\ee
This result will be very useful 
in our calculations of the CCs and the VCs. 

When $D=n$, there is no need to perform the Sommerfeld expansion 
used in the previous subsection. 
>From Eq.(\r{eh}), the total energy of the system 
can be evaluated to be 
\bea
\l{pe}
E&=&\frac{V}{\lambda^D}\int^{\infty}_{w(z^{-1})}
dw\;\frac{1}{w(1+w)}\{\mu+gkT\ln w +(1-g)kT\ln(1+w)\}\nonumber\\
&=&\frac{V}{\lambda^D}\left[\mu\ln\frac{1+w(z^{-1})}{w(z^{-1})}
-\frac{gkT}{2}(\ln\frac{1+{w(z^{-1})}}{w(z^{-1})})^2
+kT\int_{w(z^{-1})}^{\infty}dw\;\frac{\ln(1+w)}{w(1+w)}\right].
\eea
The above expression can be rewritten in a more elegant form 
by using Eq.(\r{wu}): 
\be
\l{ee}
\frac{E}{V}=\frac{1}{2}gkT\lambda^D \rho^2+
\frac{kT}{\lambda^D}\int^{\lambda^D \rho}_{0}
du\;\frac{u}{e^u-1},
\ee
where $e^u-1=1/w$.

First consider the low temperature and high density regions, 
i.e., $\lambda^D \rho \gg 1$.
Since $\lambda^D \rho \gg 1$ and thus $w(z^{-1}) \simeq 
e^{-\lambda^D\rho}\ll 1$, the integral in Eq.(\r{pe}) 
or Eq.(\r{ee}) can be evaluated up to $O(e^{-\lambda^D\rho})$ as
\be
\l{de}
E= \mu N-\frac{gkT}{2}\frac{\lambda^D}{V}N^2+
          \frac{\pi^2 kT}{6}\frac{V}{\lambda^D}+O(e^{-\lambda^D\rho}).
\ee
For $g=0$, Eq.(\r{de}) becomes 
\be
\l{e0}
E=\frac{\pi^2 kT}{6}\frac{V}{\lambda^D}+O(e^{-\lambda^D\rho})
\ee
since $\mu \simeq -kT \mbox{exp}(-\lambda^D \rho)\rightarrow 0^-$ 
as $T \rightarrow 0$. 
For $g\neq 0$, Eq.(\r{de}) becomes
\be
\l{en0}   
E=\frac{1}{2}N\epsilon_F\left[1+g\frac{\pi^2}{3}
    (\frac{kT}{\epsilon_F})^2 +O(e^{-\lambda^D\rho})\right].
\ee
Now the specific heat $C_V$ at constant volume is given by
\be
\l{sh2}
C_V/k=\left\{ \ba{ll}
                    \frac{2}{3}\frac{\pi^{\frac{D}{2}+2}}
                   {(2\pi\hbar)^D\Gamma(D/2)}\frac{NkT}{\rho}
                    +O(e^{-\lambda^D\rho}), & \mbox{if $g=0$}, \\
                    g\frac{2\pi^2}{3}\frac{E_0}{\epsilon_F^2}kT
                    + O(e^{-\lambda^D\rho}),   & \mbox{if $g\neq 0$}.
                   \ea
           \right.
\ee
The above results, of course, coincide with those of the general 
case when $D/n=1$. However, note that, as $T \rightarrow 0$, 
the chemical potential $\mu$ remains constant, i.e., 
equals to $\epsilon_F$ 
and the specific heat has only a linear correction on $T$ 
up to exponentially small corrections. This is a peculiar 
property of the energy spectrum 
with constant density of states \c{ck}. 
 
We will now investigate a systematic way of calculating the 
CCs and VCs of GES in the Boltzmann limit using 
the explicit result in Eq.(\r{wu}). 
In this region, Eq. (\r{ee}) can be explicitly 
calculated in terms of the density $\rho$; 
thus, all the VCs can be determined. 
A straightforward calculation leads to
\be
\l{eve}
\frac{P}{kT}=\rho+\frac{2g-1}{4}\lambda^D\rho^2+
\rho \sum_{l=1}^{\infty}
\frac{B_{l+1}}{(l+2)!}({\lambda^D}\rho)^{l+1},
\ee
where $B_l$ is the $l$th Bernoulli number 
($B_1=-\frac{1}{2},\; B_2=\frac{1}{6},\;
B_4=-\frac{1}{30}$, etc.) and 
$B_{2l+1}=0$ for $l\geq1$ \c{arf}. 
>From the above equation, 
the VCs in Eq.(\r{ve}) are found to be
\be
\l{evc}
\ba{l}
a_1=1,\\
a_2=\frac{2g-1}{4},\\
a_{2l+1}=\frac{B_{2l}}{(2l+1)!},\\
a_{2l+2}=0,\;\;\;\;\;\mbox{for}\;\forall l\geq 1.
\ea
\ee 
Remarkably, the VCs, except the second VC, are 
$g$-independent; moreover, all the even VCs, except 
$a_2(g)$, vanish. Of course, these properties of the VCs are 
consistent with the general properties depicted 
in the previous subsection.  

In order to calculate the CCs, $b_l(g)$, we will again use 
Eq.(\r{wu}). From Eq.(\r{wu}), we can deduce the following 
relation
\be
\l{bf}
\frac{{\tilde \rho}}{z}=\frac{{\tilde \rho} e^{\gamma {\tilde \rho}}}
{e^{{\tilde \rho}}-1}=\sum_{l=0}^{\infty}
\frac{B_{l}(\gamma)}{l!}{\tilde \rho}^l,
\ee
where ${\tilde \rho}\equiv \lambda^D \rho,\;\gamma\equiv 1-g$ and 
$B_{l}(\gamma)$ are the Bernoulli polynomials \c{arf}. 
Since $z({\tilde \rho})$ can be expressed in closed form 
in terms of Eq.(\r{bf}), 
the series expansion ${\tilde \rho}(z)$ in Eq.(\r{vn1}) 
is an inversion of the power series $z({\tilde \rho})$ \c{arf}. 
Using the method of calculus of residues, 
the CCs $b_l^{(2)}$ can be expressed as
\bea
\l{ecc}
l\cdot b_l^{(2)}&=&
\frac{1}{l!}\frac{d^{l-1}}{d{\tilde \rho}^{l-1}}
\left(\frac{{\tilde \rho}}{z}\right)^l|_{z=0}\nonumber\\
&=&\frac{1}{l!}B_{l-1}^{(l)}(\gamma l),
\eea   
where $B_{n}^{(l)}(x)$ are the generalized Bernoulli 
polynomials \c{erd} of order $l$:
\be
\l{gbp}
\ba{l}
\frac{t^l e^{xt}}{(e^t-1)^l}=\sum_{l=0}^{\infty}
B_{n}^{(l)}(x)\frac{t^n}{n!},\\
B_{n}^{(l+1)}(x)=\frac{n!}{l!}\frac{d^{l-n}}{dx^{l-n}}
(x-1)(x-2)\cdots(x-n),\;\;\;l\geq n.
\ea
\ee
Thus, the CCs $b_l^{(2)}$ are
\be
\l{cc2}
b_l^{(2)}(g)=\frac{1}{l^2}\prod_{m=1}^{l-1}\frac{\gamma l-m}{m}.
\ee

Notice that the coefficients $c_l(g)=b_l^{(0)}(g)$ 
in Eq.(\r{cc}) are independent of order $\nu$. 
Hence, we can obtain a remarkably simple result about 
the coefficients $b_l^{(\nu)}(g)$ for general $\nu$ 
from Eq.(\r{cc2}):
\be
\l{ccg}
b_l^{(\nu)}(g)=\frac{1}{l^{\nu}}\prod_{m=1}^{l-1}\frac{\gamma l-m}{m}.
\ee
>From the above equation and Eq.(\r{vcs}), the mirror symmetry of 
the CCs and the VCs in Eq.(\r{mcv}) can also be directly proved. 
Interestingly, the similar results have been noticed in the 
Calogero-Sutherland model \c{su71}, an anyon gas in the LLL \c{do94}, 
and the anyon model with linear energies \c{da96} 
in a regulating harmonic potential. 

When $D=n$, we have completely determined all the CCs and the VCs
and some lowest terms among them are
\be
\l{a15}
\ba{ll}
b_1=1,\;\;\; & a_1=1 \\
b_2=\frac{1-2g}{4},
\;\;\; & a_2=\frac{2g-1}{4} \\
b_3=\frac{1}{3^2}\prod_{m=1}^{2}(1-\frac{3}{m}g),
\;\;\; & a_3=\frac{1}{36} \\
b_4=\frac{1}{4^2}\prod_{m=1}^{3}(1-\frac{4}{m}g),
\;\;\; & a_4=0 \\
b_5=\frac{1}{5^2}\prod_{m=1}^{4}(1-\frac{5}{m}g),
\;\;\; & a_5=-\frac{1}{3600}
\ea
\ee
and $a_6=0$, {\it etc}. 
The above results exactly coincide with 
the VCs (CCs), obtained by Sen \c{se91} 
and Dasni\`eres de Veigy \c{da96}, for anyon gas. 
This result clearly shows that an anyon gas with 
a small $\alpha$ or with linear energies ($g=1-\alpha^2$)
can be well approximated by an ideal $g$-on gas. 
Thus the anyon gas confined in the LLL and 
the anyon gas with linear energies 
in a regulating harmonic potential can be regarded 
as the ideal $g$-on gas. 
Although the CCs show complicated statistics-dependences, 
the VCs are extremely simple and statistics-independent, 
except for the second VC. This implies that there are miraculous 
cancellations in the virial expansion. 

\subsection{$\mu=0$}

Let us consider the case in which
the particle number is not conserved, $\mu=0$. 
In this case, the number of particles $N$ 
itself is not a given constant, but must be determined 
from the conditions of thermal equilibrium, 
$(\partial \Omega/\partial N)_{T,V}=\mu=0$ \c{..}. 
The particle density $\rho$ and the pressure $P$ 
given by Eqs. (\r{n1}) and (\r{p1}), respectively, become 
\be
\l{0e}
\ba{l}
\rho=\frac{1}{\lambda^D}f^{(g)}_{\frac{D}{n}}(1)
\equiv\frac{1}{\lambda^D}I_{\frac{D}{n}}^g,\\
\frac{P}{kT}=\frac{1}{\lambda^D}f^{(g)}_{1+\frac{D}{n}}(1)
\equiv\frac{1}{\lambda^D}I_{1+\frac{D}{n}}^g
\ea
\ee
where the definite integrals $I_{\frac{D}{n}}^g$ 
and $I_{1+\frac{D}{n}}^g$ are pure numbers. 
This shows that the pressure $P$ and the particle 
density $\rho$ depend only on the temperature $T$. 
Actually, for a given volume, all the thermodynamic 
quantities should be determined only by the temperature, 
and their temperature dependences should be obtainable by 
simple dimensional arguments. 
For example, the energy density $E/V$ and the pressure 
are proportional to $(kT)^{1+\frac{D}{n}}$ and 
the particle density and the entropy 
density $S/V$ to $(kT)^{\frac{D}{n}}$. 
In the adiabatic expansion (or compression), 
the above results imply that 
the volume and temperature are related by $VT^{\frac{D}{n}}$={\it constant} 
and the pressure and volume by $PV^{\frac{n}{D}+1}$={\it constant} 
as in the photon statistics or an ordinary extreme relativistic gas.

When $D/n=1$, using the $w$-integration, 
the integrals $I_{1}^g$ and $I_{2}^g$ in Eq.(\r{0e}) 
can be expressed as
\be
\l{I}
\ba{l}
I_{1}^g=\frac{\xi}{g},\\
I_{2}^g=-\frac{\xi^2}{2g}+\sum_{l=1}^{\infty}
\frac{1+\xi l}{l^2}e^{-\xi l},
\ea
\ee
where $\xi=\ln(1+p)$ and $p$ is a positive root of 
the algebraic equation, $p^g(1+p)^{1-g}=1$. 
A numerical evaluation of some of the coefficients are 
as follows:
\be
\l{Iv}
\ba{llll}
I_{1}^{\frac{1}{2}}=0.9624, & I_{1}^2=0.4812,
&I_{2}^{\frac{1}{2}}=0.9870, & I_{2}^2=0.6580 \\
I_{1}^{\frac{1}{3}}=1.1467, & I_{1}^3=0.3822,
&I_{2}^{\frac{1}{3}}=1.0785, & I_{2}^3=0.5664 \\
I_{1}^{\frac{1}{4}}=1.2891, & I_{1}^4=0.3223,
&I_{2}^{\frac{1}{4}}=1.1400, & I_{2}^4=0.5050,\;\mbox{etc}.
\ea
\ee

\section{Conclusion}
We have analytically calculated some thermodynamic 
quantities of an ideal 
$g$-on gas obeying generalized exclusion statistics. 
We have shown that the specific heat of a $g$-on gas ($g \neq 0$) 
vanishes linearly in any dimension as $T \rightarrow 0$ 
when the particle number is conserved and exhibits 
an interesting dual symmetry that relates the particle-statistics 
at $g$ to the hole-statistics at $1/g$ at low temperatures.  
At the low temperatures, the thermodynamic properties at $g\neq0$ 
are similar to those of fermion due to the existence of Fermi surface, 
unlike the case of $g=0$ (bosons) for which 
the Bose-Einstein condensation can occur. 

We have found a closed form for the 
CCs $b_l(g)$ as a function of Haldane's 
statistical interaction $g$ in $D$ dimensions. 
Unlike the low temperature case, high temperature 
behaviors are bifurcated at $g=1/2$. 
The statistical interactions at $g<1/2$ are attractive (bosonic) 
while those at $g>1/2$ are repulsive (fermionic). 
The CCs $b_l(g)$ and the VCs $a_l(g)$ are 
exactly mirror symmetric ($l$=odd) or 
antisymmetric ($l$=even) about $g=1/2$. 
The case of $g=1/2$ (semions) has 
a very peculiar property, $a_{2l}(\frac{1}{2})=b_{2l}(\frac{1}{2})=0$ 
for $\forall l \in {\bf N}$. Spinon excitations in the 
Heisenberg spin chain with inverse-square exchange 
can be described by semionic 
statistics $(g=1/2)$ \c{ha91,has}, and can also be 
related to half-filling Laughlin's boson fractional 
quantum Hall states \c{kl87}. 
Thus the high temperature behaviors 
in these models may exhibit very interesting properties. 

In the case of the energy spectrum 
with constant density of states, i.e. $D=n$, 
we have obtained full exact results and have calculated 
all the CCs and the VCs of a $g$-on gas 
and we have confirmed that they exactly agree with the virial 
coefficients of anyon gas, obtained by Dasni\`eres de Veigy \c{da96}, 
for linear energies. 
Although this result implies that an anyon gas with 
linear energies can be well approximated by the ideal $g$-on gas 
and so the anyon gas of linear energy can be regarded 
as an ideal $g$-on gas, the free anyon gas cannot be generally 
regarded as an ideal $g$-on gas, e.g., in the case with 
a nonlinear energy spectrum or with strong coupling \c{do92} 
for which a statistical dependence of the third VC 
has been shown to appear in the second order correction. 
Even so, there may still remain the nonperturbative similarities, 
such as the mirror symmetry about semions, 
between anyon statistics and GES. Note that it was reported 
by Sen \c{se92} that the third VC of anyons exhibits similar mirror 
symmetry about semions. It will be very interesting if 
the higher virial coefficients of an anyon gas 
turn out to be also mirror symmetric or antisymmetric 
about semions. 

In the case of $\mu=0$, the number of particles $N$ is not 
a given constant as in a degenerate gas but the quantity 
being determined from the conditions of thermal equilibrium. 
As in the photon statistics, 
for a given volume, all the themodynamic 
quantities of a $g$-on gas depend only on the temperature and 
these dependences can be determined by simple 
dimensional arguments. 

In their paper \c{nw94}, Nayak and Wilczek suggested that the 
electrons in Mott insulators behave as $2$-ons 
and, thus, a Mott insulator has a Fermi surface of 
anomalous size and a specific heat of anomalous value. 
Our results show that the specific heat and the volume of the 
Fermi surface of $2$-ons are twice as large as ordinary fermions 
in the units of the Fermi energy. Also, Ihm suggested \c{ihm} that 
the reduction of the number of allowed configurations of 
crystalline ice can be recognized 
as arising from the fractional exclusion 
due to the so-called ice rule and that this reduction can be 
described as the GES with the statistical interaction $g=0.867$. 
However, it still remains open to construct an appropriate 
three-dimensionsional model describing the GES, 
unlike the one-dimensional soluble models nicely fitting with 
the notion of the GES such as Calogero-Sutherland model \c{is94}, 
antiferromagnetic spin chain with 
inverse-square exchange \c{has}, and the
twisted Haldane-Shastry model \c{fk96}. 
Hence, it will be very interesting although very difficult 
to construct higher 
dimensional models nicely fitting with the notion of GES.

\section*{ACKNOWLEDGEMENTS}
This work was supported by the Korean Science and 
Engineering Foundation 
(94-1400-04-01-3 and the Center for Theoretical Physics) and 
by the Korean Ministry of Education (BSRI-95-2441).
\\
{\it Note Added.}: After the completion of the paper, we 
noticed Ref. \c{iamp}, 
with which some parts of our main results overlap.

\end{document}